\documentclass[aps,prc,groupedaddress,showpacs,manuscript]{revtex4-1}
\usepackage{graphicx}
\usepackage{colordvi}

\begin{document}

\title{$^3$H/$^3$He ratio as a probe of the nuclear symmetry energy at sub-saturation densities}
\author {Yongjia Wang$\, ^{1,2}$,
Chenchen Guo$\, ^{1,3}$,
Qingfeng Li$\, ^{1}$\footnote{E-mail address: liqf@hutc.zj.cn},
and
Hongfei Zhang$\, ^{2}$
}

\affiliation{
1) School of Science, Huzhou University, Huzhou 313000, P.R. China \\
2) School of Nuclear Science and Technology, Lanzhou University, Lanzhou 730000, P.R. China  \\
3) College of Nuclear Science and Technology, Beijing Normal University, Beijing 100875, P.R. China \\
 }
\date{\today}

\begin{abstract}
Within the newly updated version of the Ultra-relativistic quantum molecular dynamics (UrQMD) model in which the Skyrme potential energy-density functional is introduced, the yield ratio between $^3$H and $^3$He clusters emitted from central $^{40}$Ca+$^{40}$Ca, $^{96}$Zr+$^{96}$Zr, $^{96}$Ru+$^{96}$Ru, and $^{197}$Au+$^{197}$Au collisions in the beam energy range from 0.12 to 1 GeV$/$nucleon is studied. The recent FOPI data for the $^3$H$/$$^3$He ratio are compared with UrQMD calculations using 13 Skyrme interactions (all exhibiting similar values of iso-scalar incompressibility but very different density dependences of the symmetry energy).
It is found that the $^3$H$/$$^3$He ratio is sensitive to the nuclear symmetry energy at sub-saturation densities. Model calculations with moderately soft
to linear symmetry energies are in agreement with the experimental FOPI data.
This result is in line with both, the recent constraints on the low-density symmetry energy available in the literature and our previous results for the high-density symmetry energy obtained with the elliptic flow of free nucleons and hydrogen isotopes as a sensitive probe.

\end{abstract}


\maketitle
\section{Introduction}
The density dependence of the nuclear symmetry energy is a hot topic in both nuclear and astrophysics, due to its
importance for the structure of exotic nuclei, the dynamics of heavy-ion collisions (HICs) induced by neutron-rich nuclei, and the properties of neutron stars and other astrophysical phenomena. It is also one of the important goals of the
current and future rare isotope beam facilities (e.g. CSR at HIRFL, FAIR at GSI, SPIRAL2 at GANIL, FRIB at MSU, RIBF at RIKEN) around the world. For recent reviews see Refs.~\cite{BALi08,Horowitz:2014bja,Li:2014oda}.

The energy per nucleon of isospin asymmetric nuclear matter can be generally expressed as $e(\rho,\delta)=e_{0}(\rho,0)+e_{\rm sym}(\rho)\delta^{2}$,
where $\delta=(\rho_{n}-\rho_{p})/(\rho_{n}+\rho_{p})$ is the
isospin asymmetry, and $\rho_n$, $\rho_p$, and $\rho$ are the neutron, proton and total nucleon densities.
$e_{0}(\rho,0)$ is the energy per nucleon of the isospin symmetric nuclear matter, while $e_{\rm sym}(\rho)$ is the nuclear symmetry energy.
 Thanks to the continuing endeavor
of both nuclear physicists and astrophysicists in recent years, many sensitive probes from nuclear structure, nuclear reactions and
 neutron stars have been used to estimate parameters (e.g., the coefficient $S_0=e_{\rm sym}(\rho_{0})$ and the slope parameter $L=3\rho_{0}\left(\frac{\partial{e_{\rm sym}(\rho)}}{\partial\rho}\right)|_{\rho=\rho_{0}}$)
 of the symmetry energy at saturation density ($\rho_{0}$). So far, the values of the nuclear symmetry energy at $\rho_0$ and at $\rho\approx 0.11 fm^{-3}$ have been relatively well constrained but its value at other densities or, generally, its density dependence has still large uncertainties (see, e.g., Refs.~\cite{Tsang:2012se,Lattimer:2012xj,RocaMaza:2012mh,Zhang:2013wna,brown,Danielewicz:2013upa,Fan:2014rha}).

The ratio of $^3$H and $^3$He yields emitted from HICs has been suggested as a sensitive probe of the nuclear symmetry energy within both the Boltzmann-Uehling-Uhlenbeck (BUU) type
and the quantum molecular dynamics (QMD) type transport models \cite{Chen:2003qj,Chen:2004kj,Li:2005kqa,Zhang:2005sm}, still with some puzzling inconsistency.
For example, in Refs.~\cite{Li:2005kqa,Zhang:2005sm} two QMD type model calculations showed that the yield of $^3$H calculated with a soft symmetry energy is larger than that
with a stiff one, while in Ref.~\cite{Chen:2003qj} the isospin-dependent BUU (IBUU) model calculations showed the opposite trend.
In Ref.~\cite{Chen:2004kj}, using the Gogny effective interaction (MDI) in the IBUU model, it was found that both, $^3$H and $^3$He yields, did not exhibit significant differences
between the results for a soft and a stiff symmetry energy. Moreover, it was shown in Ref.~\cite{Yong:2009te} that, using the IBUU04 model incorporated with a phase-space coalescence
afterburner, the $^3$H/$^3$He ratio was found not to be sensitive to the nuclear symmetry energy any more. Furthermore, a recent study in Ref.~\cite{Youngs} showed a large discrepancy
regarding the $^3$H$/$$^3$He ratio at low kinetic energies between Michigan State University experimental data and a BUU (often called pBUU) model  as well as an improved QMD (often called ImQMD) model simulations,
regardless of which parameterized symmetry energy is employed.
Thus, in view of the current status for the detection of both $^3$H and $^3$He clusters, the sensitivity of the nuclear symmetry energy to the $^3$H$/$$^3$He ratio is a subject of continued interest.

Recently, the large-acceptance apparatus FOPI at the Schwerionen-Synchrotron (SIS) at GSI has been used to collect a large amount of yield data
for light charged particles (protons, $^2$H, $^3$H, $^3$He, and
$^4$He) from intermediate energy HICs which has been made available in Refs.~\cite{FOPI:2010aa,FOPI:2011aa}. This data set offers new opportunities for studying the $^3$H$/$$^3$He ratio over wide ranges of both, beam energy and system size. Moreover, by using the updated Ultrarelativistic Quantum Molecular Dynamics (UrQMD) model in which the Skyrme potential energy density functional is introduced, the newly measured flow data of light charged particles can be reproduced quite well \cite{wyj,wyj-sym}. In this version of the UrQMD model, the stiffness of the symmetry energy can be more
consistently selected within a broad range by choosing different
Skyrme interactions, rather than by varying the strength parameter $\gamma$ in
the potential term of the symmetry energy which, in addition,
cannot be used to express a very soft symmetry energy~\cite{Dong:2012zza}.
In view of these developments, it seems timely to re-examine the sensitivity of the $^3$H/$^3$He ratio to the symmetry energy and to see whether it can provide firm constraints on the stiffness of the density-dependent symmetry energy.

The paper is arranged as follows. In the next section the new version of UrQMD with the use of Skyrme potential energy density functionals and its key parametrizations are presented. In Sec. III, results for $^3$H, $^3$He and their ratio from HICs at SIS/GSI energies are shown and discussed. Finally, a summary is given in Sec. IV.

\section{Model descriptions}
The UrQMD model has been widely and successfully used to study nuclear reactions of \emph{p}+\emph{p}, \emph{p}+\emph{A} and \emph{A}+\emph{A} systems within a
large range of beam energies, from low SIS/GSI up to the LHC/CERN \cite{Li:2011zzp,Bass98,Bleicher:1999xi,Li:2012ta}. In the present code, the nuclear effective interaction potential energy $U$ is derived from the integration of the Skyrme potential energy density functional, $U_{\rho}=\int u_{\rho}d^3r$, and $u_{\rho}$ reads
\begin{eqnarray}
u_{\rho}&=&\frac{\alpha}{2}\frac{\rho^2}{\rho_0}+
\frac{\beta}{\eta+1}\frac{\rho^{\eta+1}}{\rho_0^{\eta}}+
\frac{g_{sur}}{2\rho_0}(\nabla\rho)^2\nonumber\\
&&
+\frac{g_{sur,iso}}{2\rho_0}[\nabla(\rho_n-\rho_p)]^2
+(A\rho^{2}+B\rho^{\eta+1}+C\rho^{8/3})\delta^2\nonumber\\
&&
+g_{\rho\tau}\frac{\rho^{8/3}}{\rho_0^{5/3}}.
\end{eqnarray}
Here, $\alpha$, $\beta$, $\eta$, $g_{sur}$, $g_{sur,iso}$, \emph{A}, \emph{B},\emph{ C}, and $g_{\rho\tau}$ are parameters which can be directly calculated using Skyrme parameters (see, e.g., Refs.~\cite{wyj,Zhang:2006vb}).
In this work, we choose 13 Skyrme interactions Skz4, Skz2, SV-mas08, SLy4, MSL0, SkO', SV-sym34, Rs, Gs, Ska35s25, SkI2, SkI5, and SkI1\cite{Dutra:2012mb}, which give quite similar values of the incompressibility $K_0$ but different \emph{L} values (the saturation properties of selected forces
are shown in Table~\ref{skyrme}). In addition, the symmetry energies at $\rho$=0.08, 0.055, and 0.03 $fm^{-3}$ are also shown in Table~\ref{skyrme}.

\begin{table*}[htbp]
\caption{\label{tab:table1} Properties of nuclear matter at various densities as calculated by selected Skyrme parametrizations used in this work. All entries are in MeV, except for density in $fm^{-3}$. }
\begin{tabular}{ccccccccccccccc}
\hline
\hline
&
&&
&&
&&
&&$\rho=0.08$
&&$\rho=0.055$
&&$\rho=0.03$
\\
&$\rho_{0}$ &&$K_0$ &&$S_0$ &&$L $  &&$e_{sym}(\rho)$ &&$e_{sym}(\rho)$&&$e_{sym}(\rho)$\\
\hline
Skz4 &0.160  &&230.08 &&32.01&&5.75&&26.73 &&22.48&&15.91\\
Skz2 &0.160 &&230.07&&32.01&&16.81&&24.90&&20.30&&13.82\\
SV-mas08&0.160 &&233.13&&30.00&&40.15&&20.53&&16.06&&10.44\\
SLy4&0.160&&229.91&&32.00&&45.94&&22.17&&17.68&&11.89\\
MSL0   &0.160&&230.00&&30.00&&60.00&&18.39&&13.90&&8.72\\
SkO'&0.160 &&222.36&&31.95&&68.94&&19.12&&14.36&&8.97\\
SV-sym34 &0.159&&234.07&&34.00&&80.95&&19.38&&14.15&&8.50\\
Rs&0.158&&237.42&&30.82&&86.39&&16.48&&11.76&&6.92\\
Gs&0.158&&237.29&&31.13&&93.31&&16.16&&11.34&&6.53\\
Ska35s25 &0.158&&241.30&&36.98&&98.89&&20.39&&14.73&&8.78\\
SkI2  &0.158&&240.93&&33.37&&104.33&&17.15&&12.21&&7.26\\
SkI5 &0.156&&255.79&&36.64&&129.33&&17.75&&12.34&&7.22\\
SkI1 &0.160&&242.75&&37.53&&161.05&&14.21&&8.69&&4.15\\
\hline
\hline
\end{tabular}
\label{skyrme}
\end{table*}

\begin{figure}[htbp]
\centering
\includegraphics[angle=0,width=0.8\textwidth]{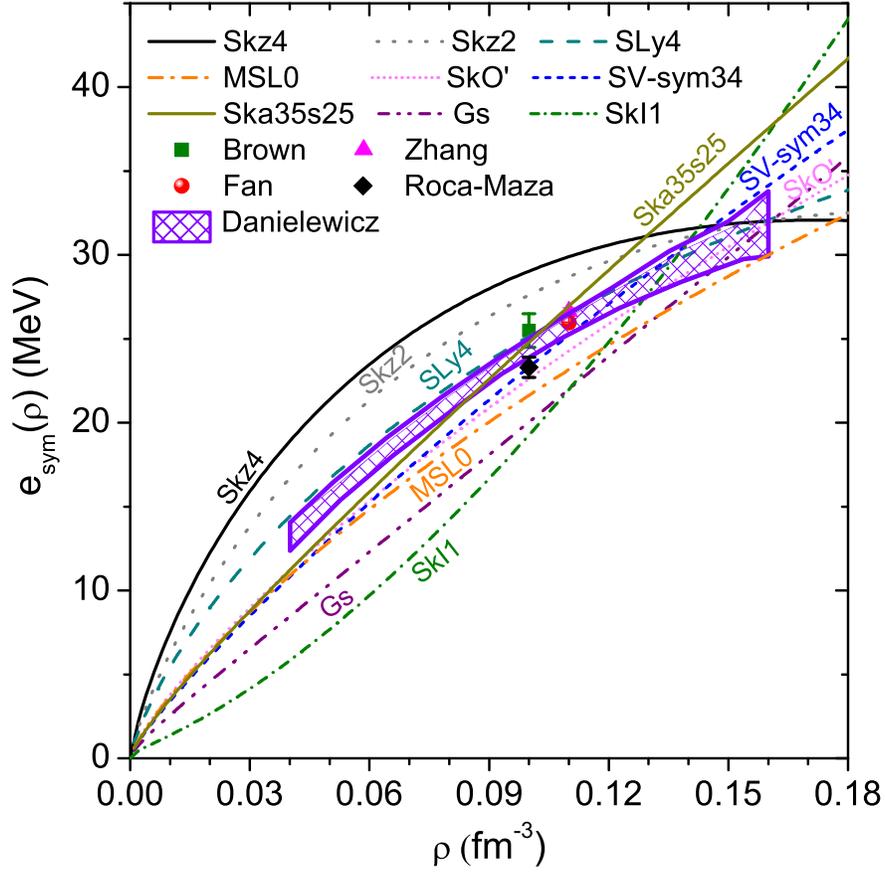}
\caption{\label{fig1}(Color online) Density dependence of the symmetry energy for Skz4, Skz2, SLy4, MSL0, SkO', SV-sym34, Ska35s25, Gs, and SkI1.
The shaded region exhibits the result obtained by Danielewicz \emph{et al.} \cite{Danielewicz:2013upa}.
Four different scattered symbols represent recent constraints obtained by Roca-Maza \emph{et al.} \cite{RocaMaza:2012mh}, Brown\cite{brown},
Zhang \emph{et al.} \cite{Zhang:2013wna} , and Fan \emph{et al.} \cite{Fan:2014rha}, respectively. }
\end{figure}

Figure~\ref{fig1} shows the density dependence of the symmetry energy for Skyrme interactions Skz4, Skz2, SLy4, MSL0, SkO', SV-sym34, Ska35s25, Gs, and SkI1. For comparison, very recent constraints extracted from nuclear properties as, e.g., binding energy, neutron skin thickness, isovector giant quadrupole resonance and isobaric analog states~\cite{RocaMaza:2012mh,Zhang:2013wna,brown,Danielewicz:2013upa,Fan:2014rha}, are also presented. The symmetry energy predicted by SLy4 lies quite close to the upper limit of the result obtained by Danielewicz \emph{et al.}~\cite{Danielewicz:2013upa}, and also covers the results obtained by Brown~\cite{brown} and by Zhang and Chen~\cite{Zhang:2013wna}. The symmetry energy determined by Roca-Maza \emph{et al.}~\cite{RocaMaza:2012mh} for the density 0.1 fm$^{-3}$ is very close to the results predicted with SkO' as well as SV-sym34. Furthermore, the density dependent symmetry energies in MSL0, SkO', SV-sym34, and Ska35s25 are very close to each other at low densities (below 0.06 fm$^{-3}$) but well separated at high densities.

The treatment of the collision term is the same as in our previous work in which the FP4 parametrization of
the in-medium nucleon-nucleon cross section is employed~\cite{wyj}. Several ten thousand events of HICs are simulated in order to achieve small enough statistical uncertainties; error bars are hence not shown in the following figures, also for the sake of clarity. The simulation is stopped at 150 fm/c and an isospin-dependent minimum span tree algorithm (iso-MST) is used to construct clusters.
Proton-proton or neutron-neutron (proton) pairs with relative distances smaller than $R_0^{pp}=2.8$ fm or $R_0^{nn}=R_0^{np}=3.8$ fm, respectively, and relative
momenta smaller than $P_0 =0.25$ GeV/$c$ are considered to belong to the same
cluster.

\section{Results}
\begin{figure}[htbp]
\centering
\includegraphics[angle=0,width=0.8\textwidth]{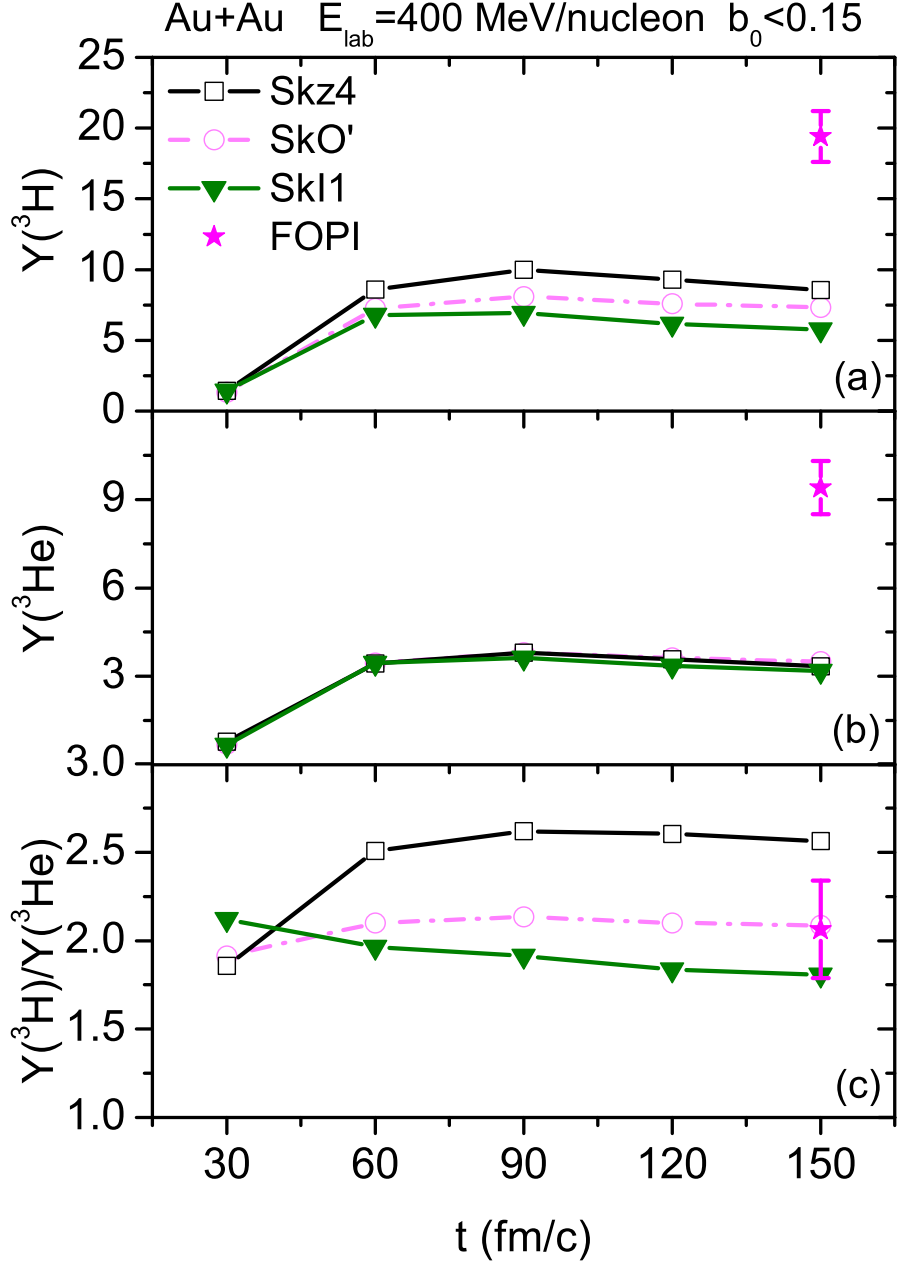}
\caption{\label{fig2}(Color online) Time evolution of $^3$H [in plot (a)] and $^3$He [in (b)] yields and their ratio $^3$H/$^3$He [in (c)] from Au+Au collisions at $E_{lab}$=400 MeV$/$nucleon with centrality $b_0<0.15$. Calculations (lines with symbols) with Skz4, SkO', and SkI1 are compared with the 4$\pi$ reconstructed FOPI data \cite{FOPI:2010aa} (solid stars). It is noticed that the magnitude of experimental errors for the $^3$H/$^3$He ratio is calculated using the propagation of errors, assuming the measurements of $^3$H and $^3$He yields are uncorrelated, while this assumption is not necessarily valid due to possible correlations in the detection system.}
\end{figure}
Figure~\ref{fig2} illustrates the time evolution of the yields of $^3$H (a), $^3$He (b), and of their ratio $^3$H/$^3$He (c) from Au+Au central ($b_0<0.15$) collisions at $E_{\rm lab}$=400 MeV$/$nucleon. The reduced impact parameter $b_0$ is defined
as $b_0=b/b_{\rm max}$ with $b_{\rm max} = 1.15 (A_{P}^{1/3} + A_{T}^{1/3})$. The Skz4, SkO', and SkI1 forces are chosen for this illustration as they correspond to the top, middle, and bottom curves at low densities among the cases shown in Fig.~\ref{fig1}. From Fig.~\ref{fig2} it is, first of all, seen that most of the $^3$H and $^3$He clusters are produced within the time span from 30~fm$/$c to 60~fm$/$c. During this interval, the central nucleon density drops from normal to sub-normal (see, e.g., Fig. 2 in Ref.~\cite{Li:2005kqa} which shows that the high-density state is limited to the time span from about 5~fm$/$c to 30~fm$/$c for a similar colliding system).

Secondly, the UrQMD model calculations underestimate the yields of both $^3$H and $^3$He. Due to the lack of spin degrees of freedom
in the QMD-like models, this problem is generic and can not be resolved by considering uncertainties in the stiffness of the equation of state or in the medium modification of
the two-body collision term. However, one sees that the $^3$H yield is sensitive to the Skyrme interaction whereas the $^3$He yield does not exhibit this sensitivity.
As evident from Table~\ref{skyrme}, these Skyrme parametrizations are selected to have large differences only in the value of $L$.
However, the emission of $^3$H is affected by the nuclear symmetry potential, while the $^3$He yield is also affected by the Coulomb potential between two protons which reduces
the sensitivity to the symmetry energy, as discussed in Ref.~\cite{Guo:2012aa}.
Finally, we see that the SkI1 parametrization results in the smallest $^3$H yield, while Skz4 leads to the largest $^3$H yield.
This follows from the fact that, in a neutron-rich environment at sub-normal density, Skz4 gives a more repulsive symmetry potential,
leading to a larger phase-space distribution for neutrons.
Hence, more neutrons and neutron-rich light clusters such as $^3$H are produced. Similar results were also obtained in two other QMD-type models \cite{Li:2005kqa,Zhang:2005sm}.

If we turn to Fig.~\ref{fig2}(c), we observe that, during the initial stage at times $t<$40~fm/c, the SkI1 force gives the larger $^3$H/$^3$He ratio while, at the final stage, this is the case for Skz4. At $t<$40 fm/c, $^3$H and $^3$He consist of protons and neutrons which evolve mainly from the supra-saturation density region and are emitted early; a stiff symmetry energy will thus cause more neutrons and less protons to be emitted than a soft one. More $^3$H can be formed, resulting in a higher value of the $^3$H/$^3$He ratio as well as of the ratio of free neutrons to protons. Very similar results can be found in other QMD-type model calculations \cite{Li:2005kqa,Kumar:2011td}. As the reaction proceeds, the formed hot and dense system
decompresses and more and more nucleons and light clusters will evolve and emerge from a low-density environment. The opposite effect of the symmetry energy on the $^3$H/$^3$He ratio should appear. It follows that, at the asymptotic stage, the $^3$H/$^3$He ratio predominantly reflects the symmetry energy at sub-saturation densities.

\begin{figure}[htbp]
\centering
\includegraphics[angle=0,width=0.9\textwidth]{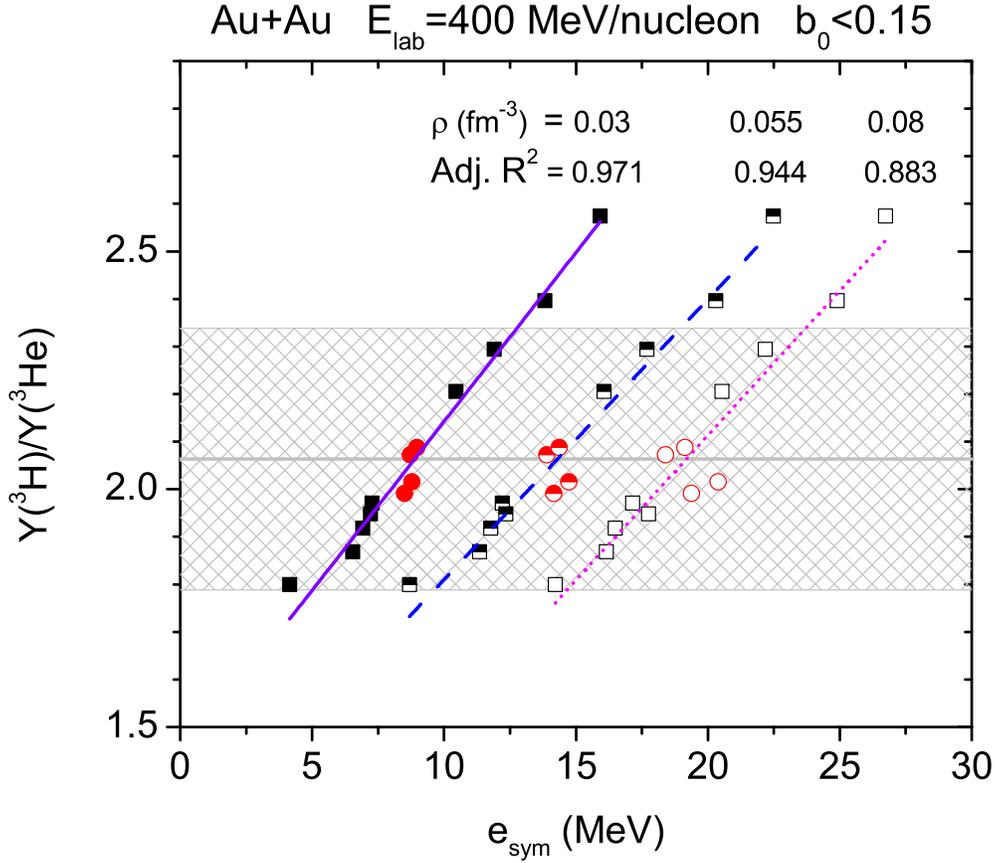}
\caption{\label{fig3}(Color online) $^3$H/$^3$He ratio as a function of symmetry energy at densities of $\rho$=0.08 (open symbols), 0.055 (half-solid symbols), and 0.03 (solid symbols) $fm^{-3}$. Four circle symbols in each bunch denote calculations with MSL0, SkO', SV-sym34, and Ska35s25, and square symbols represent calculations with other 9 Skyrme interactions listed in Table~\ref{skyrme}. The lines represent linear fits to calculations for the three density cases. Correspondingly, the Adj. $R^2$ values are also given. The shaded region indicates the FOPI data of $^3$H/$^3$He ratio \cite{FOPI:2010aa}.}
\end{figure}

Figure~\ref{fig3} shows the $^3$H/$^3$He ratios calculated with the 13 selected Skyrme parametrizations as a function of the symmetry energy at three sub-normal density points: 0.08, 0.055, and 0.03 $fm^{-3}$. The line in each bunch represents a linear fit to the calculations, the respective value of the adjusted coefficient of determination (Adj. $R^2$) is also shown. It can be seen that the linearity between the $^3$H/$^3$He ratio and the symmetry energy increases with decreasing density, which indicates a strong correlation between them at low densities. The $^3$H/$^3$He ratios calculated with MSL0, SkO', SV-sym34, and Ska35s25 (which give almost the same value of the symmetry energy at $\rho$=0.03 $fm^{-3}$, see Table~\ref{tab:table1}) are close to each other and centered in the shaded band, while the results obtained with Skz4 and Skz2 fall outside the band. Obviously, the large uncertainty of the experimental data prevents us from getting a tighter constraint on the density-dependent symmetry energy. Furthermore, it is noted \cite{Li:2008fn,Zhang:2012qm} that different cluster recognition criteria affect the neutron$/$proton as well as $^3$H/$^3$He ratios, especially at small kinetic energies. Therefore, the comparison of the full energy spectrum with the experimental data seems mandatory for a future study.

\begin{figure}[htbp]
\centering
\includegraphics[angle=0,width=0.8\textwidth]{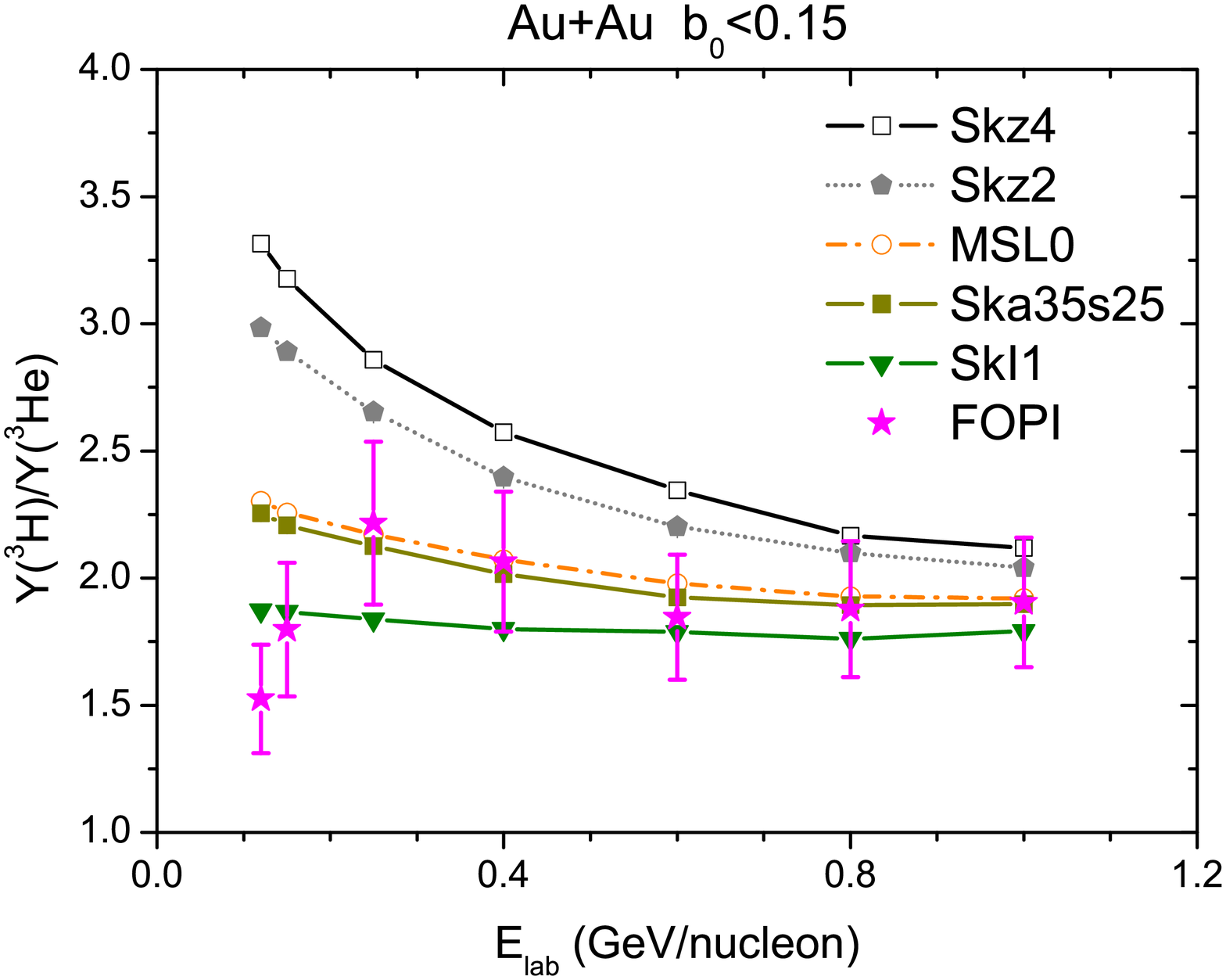}
\caption{\label{fig4}(Color online) Excitation function of the $^3$H/$^3$He ratio from central ($b_0<0.15$) Au+Au collisions. Calculations with Skz4, Skz2, MSL0, Ska35s25, and SkI1 are represented by different lines with symbols. The FOPI data (stars) are taken from Ref.~\cite{FOPI:2010aa}.}
\end{figure}

Luckily, the comparison to experimental $^3$H/$^3$He data as functions of beam energy and system size and composition has become possible, supplying a more systematic and thus more consistent information on the symmetry energy. Figure~\ref{fig4} displays firstly the $^3$H/$^3$He ratio as a function of beam energy. Calculations performed with Skz4, Skz2, MSL0, Ska35s25, and SkI1 are compared to the experimental data represented by the stars.
At low beam energies (below 200 MeV$/$nucleon), the ratio is quite sensitive to the density dependence of the symmetry energy, however, the experimental data cannot be well reproduced. The disagreement of calculations with data implies again that the method for constructing clusters is not fully valid at low beam energies. Above 200 MeV$/$nucleon, the ratio is still sensitive to the symmetry energy, but the sensitivity decreases with increasing beam energy due to the increase of both, the nucleon density and the number of nucleon-nucleon collisions. Furthermore, the calculations with MSL0 and Ska35s25, for which the difference in $L$ is as large as $\sim$39 MeV, are very close to each other, indicating the sensitivity of the $^3$H/$^3$He ratio to the stiffness of the symmetry energy is more obvious at low densities.

\begin{figure}[htbp]
\centering
\includegraphics[angle=0,width=0.8\textwidth]{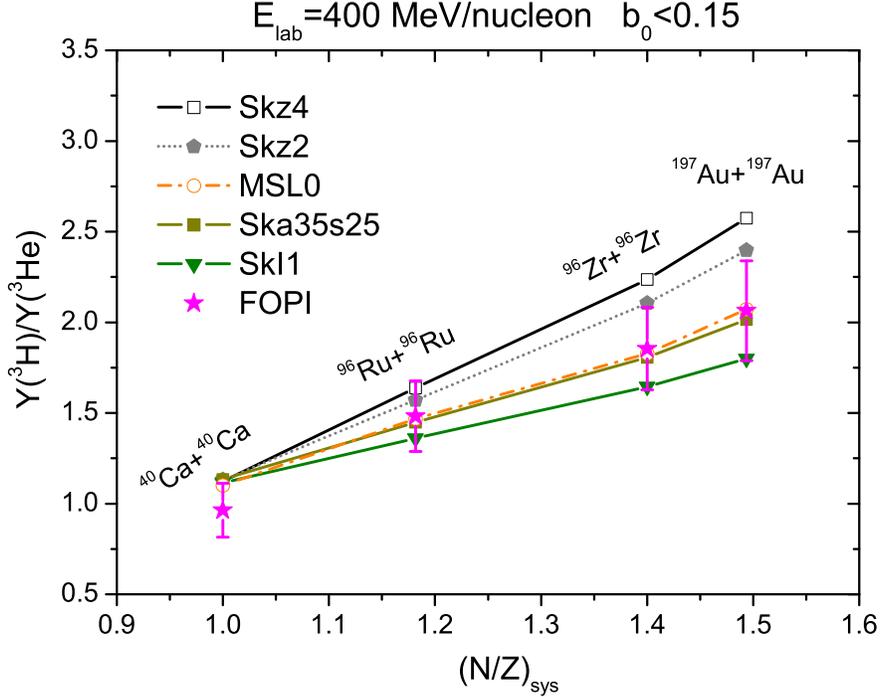}
\caption{\label{fig5}(Color online)  $^3$H/$^3$He ratio from central ($b_0<0.15$) $^{40}$Ca+$^{40}$Ca, $^{96}$Ru+$^{96}$Ru, $^{96}$Zr+$^{96}$Zr, and $^{197}$Au+$^{197}$Au collisions at $E_{lab}$=400 MeV$/$nucleon as a function of neutron/proton ratio of the colliding system. The FOPI data (stars) are taken from Ref.~\cite{FOPI:2010aa}.}
\end{figure}

Figure~\ref{fig5} displays the calculated $^3$H/$^3$He ratios as a function of the neutron/proton ratio of the colliding systems in comparison with the FOPI data for central collisions at $E_{\rm lab}$=400 MeV$/$nucleon. The reaction systems are $^{40}$Ca+$^{40}$Ca, $^{96}$Ru+$^{96}$Ru, $^{96}$Zr+$^{96}$Zr, and $^{197}$Au+$^{197}$Au. For the isospin symmetric $^{40}$Ca+$^{40}$Ca case, the results with the five selected Skyrme forces are very close to each other. With the increase of isospin asymmetry by varying systems from $^{96}$Ru+$^{96}$Ru to $^{197}$Au+$^{197}$Au, the calculated results are well separated due to an increasingly stronger effect of the symmetry energy at sub-normal densities. Especially, calculations using both MSL0 and Ska35s25 (as well as SkO' and SV-sym34, not shown in the figure), which represent a moderately soft to linear symmetry energy, reproduce the data fairly well. Although a desirable tighter constraint to the density dependence of the symmetry energy is still not achieved here, partly due to the large experimental uncertainties, a very satisfactory consistency among the presented comparisons is achieved.

Utilizing the recently updated UrQMD model in which the Skyrme potential energy density functional is adopted, we have studied the isospin pair $^3$H and $^3$He production in heavy-ion collisions at intermediate energies. The $^3$H/$^3$He yield ratio is shown to exhibit a large sensitivity to the nuclear symmetry energy at sub-saturation densities. This result is similar to most previous studies using BUU-type and QMD-type transport models, but differences in the influence of the symmetry energy on the yield of (and ratio between) light clusters are also observed between these model calculations, which certainly requires further investigations.

In the current work, simulations with 13 selected Skyrme interactions are compared with the recent FOPI data. It is found that those calculations for the dependence on the nucleon-density, beam energy, and system size and composition of the $^3$H/$^3$He ratio employing MSL0, SkO', SV-sym34, and Ska35s25 (which parameterize a moderately soft to linear symmetry energy) are all in good semi-quantitative agreement with FOPI data. Calculations using Skz4 and Skz2 which parameterize a very soft symmetry energy are far from the data. It is observed that the current extraction of the stiffness of the symmetry energy at sub-normal densities with the $^3$H/$^3$He ratio as a probe is in line with previous studies for the low-density symmetry energy. It is more exciting that this result is also consistent with previous results based on the elliptic flow of free nucleons (and hydrogen isotopes) as a probe~\cite{wyj-sym,Russotto:2011hq,Cozma:2013sja} which  mainly provide information on the symmetry energy at supra-normal densities.

\begin{acknowledgements}
We thank Professor W. Trautmann for a careful reading of
the manuscript. We acknowledge support by the computing server C3S2 in Huzhou University. The work is supported in part by the National Natural
Science Foundation of China (Nos. 11175074 and 11375062) and the project sponsored by SRF for ROCS, SEM.
\end{acknowledgements}

\end{document}